\documentstyle[aps,prl,epsf]{revtex}
\def\Dagcomp{^{\vphantom{\dagger}}}

\def\bsqre{\vrule width1.4ex height1.4ex depth 0ex}
\begin{document}
\draft
\tighten
\twocolumn[\hsize\textwidth\columnwidth\hsize\csname
@twocolumnfalse\endcsname
\title{Optical properties of a Quantum-Dot Cascade Structure}
\author{V.M. Apalkov and Tapash Chakraborty$^*$}
\address{Max-Planck-Institut f\"ur Physik Komplexer Systeme, 
01187 Dresden, Germany}
\date{\today}
\maketitle
\begin{abstract}
We report on our theoretical studies of the luminescence spectra
of a quantum cascade laser where the quantum wells in the active
regions are replaced by parabolic quantum dots. We analyze the 
influence of shape and size of the dots on the luminescence
spectra. The emission spectra have interaction induced blueshift
which increases almost linearly with increasing electron number.
The blueshift is smaller for larger and  non-circular dots.
For large dots, shape of the emission line has weak dependence 
on the shape of quantum dots. 

\end{abstract}
\vskip2pc]
\narrowtext
Ever since the original work on quantum cascade laser (QCL) 
by Faist et al. \cite{first,vertical,vertprl} in 1994, the unipolar 
semiconductor laser based on intersubband transitions in 
coupled quantum wells has undergone rapid developments. QCLs 
created in InGaAs/AlInAs systems have achieved record high power 
outputs in the mid-infrared range that has the potential for 
wide-ranging applications \cite{apply,sense,sky}. Unlike in 
conventional lasers where it is fixed by the energy bandgap of 
the material, emission wavelength in a QC laser is essentially 
determined by the thickness of the active layers. Therefore,
a suitable tailoring of the active layer thickness can produce
a laser wavelength over a wide range (3-12$\mu$m) of the 
IR spectrum for the same material. In addition, QC lasers 
have much higher power than diode lasers because of the 
cascade scheme. In this scheme, an electron, after it has 
emitted a laser photon in the first active region of the device, 
is reinjected into the next stage that follows where it emits 
another photon. In this way an injected electron creates, in 
principle, $N_{st}$ laser photons as it traverses the device. 
Here $N_{st}$ is the number of stages. Application of an external 
perpendicular magnetic field on the QCL has uncovered interesting 
quantum oscillations in magnetotransport \cite{leotin},
and in future might provide magnetic tuning \cite{qclmag} of the
laser wavelength for a given QC laser.

In this letter, we report on the physics of a quantum cascade 
structure where the quantum wells in the active regions are 
replaced by parabolic (symmetrical and asymmetrical)
quantum dots (QD). The latter objects, popularly known as artificial 
atoms \cite{chakmak}, where the electron motion is quantized in all 
three spatial directions, have been receiving much attention. 
These zero-dimensional quantum confined systems are useful 
for investigating the fundamental concepts of nanostructures 
\cite{chakmak,qdbook,tarucha} as well as for its vast application 
potentials. Important recent observations in QD transport 
measurements are the shell filling, half-filling dictated by Hund's 
rule \cite{tarucha,shell}, for circular as well as 
elliptical quantum dots \cite{madhav,austing}. 
On the other hand, in recent years there has been considerable 
progress in quantum-dot laser research \cite{qdrev}. Because of 
their discrete atom-like states, quantum-dot lasers are expected 
to have better performance than the quantum-well lasers 
\cite{sakaki}. A quantum-dot cascade laser (QDCL) \cite{qdcl}
with rectangular confinement potentials in the electron plane
is predicted to exhibit a large blueshift ($\sim55$ meV) in diagonal
transitions (for a dot size of 10 nm$\times$10 nm) when the 
electron number was increased from 1 to 6 in the dot. Most of 
the quantum dot studies to date, are however, for a parabolic 
confinement potential (isotropic as well as anisotropic) that 
turned out to be more appropriate for observing shell structure 
and the influence of shape and size of the dot on the electronic 
states. QD cascade lasers, although not yet realized, might be
more efficient than the present-day quantum-well cascade lasers
\cite{efficient}. Knowledge of its physical properties might 
be useful in future development of this nanostructured light
source.

\begin{figure}
\centerline{
\epsfxsize=2.6in
\epsfbox{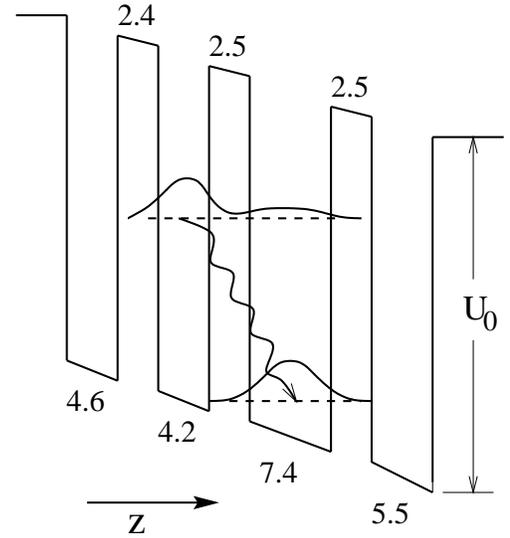}}
\vspace*{0.1in}
\protect\caption[sample]
{\sloppy{Energy band diagram (schematic) of the active region 
of a quantum cascade laser structure under an average applied 
electric field of 55 kV/cm \protect\cite{blaser}. Only one period 
of the device is shown here. The relevant wave functions (moduli 
squared) as well as the transition corresponding to the laser 
action are also shown schematically. The numbers (in nm) are the 
well (Ga$_{0.47}$In$_{0.53}$As) and barrier (Al$_{0.48}$In$_{0.52}$As) 
widths. Material parameters considered in this work are: electron
effective mass $m_e^*$ (Ga$_{0.47}$In$_{0.53}$As)=0.043 $m_0$,
$m_e^*$ (Al$_{0.48}$In$_{0.52}$As)=0.078 $m_0$, the conduction band
discontinuity, $U_0=520$ meV, the nonparabolicity coefficient,
$\gamma_w=1.3\times10^{-18}$ m$^2$ for the well and
$\gamma_b=0.39\times10^{-18}$ m$^2$ for the barrier.
}}
\label{figsample}
\end{figure}

The single-electron Hamiltonian for our system is
$${\cal H}'=\frac{p_x^2}{2m^*}+\frac{p_y^2}{2m^*}+V_{\rm plane}
(x,y) +\frac{p_z^2}{2m^*}+V_{\rm conf}(z)$$
where the confinement potential in the $z$-direction is
$$V_{\rm conf}(z)=-eFz + \left\{
\begin{array}{ll}
0 & \mbox{for wells} \\
U_0 & \mbox{for barriers}
\end{array}
\right.
$$
with $F$ being the electric field in the $z$-direction, 
$m^*$ is energy dependent due to band non-parabolicity,
$m^*=m^*_e\left(1+2Em^*_e\gamma/\hbar^2\right)$, and
$U_0$ is the conduction band discontinuity \cite{qdcl}. All 
material parameters of our model of QDCL are given in
Fig.~\ref{figsample}, where we show one active region of the 
QC laser structure reported by Blaser et al. \cite{blaser}. 
The structure emits at a wavelength of 10.5 $\mu$m,
depending upon the electric field and temperature \cite{blaser}.
For our present work, details of the laser structure are
not important and the most relevant system for us is the 
active region of Fig.~\ref{figsample}. The confinement potential 
in the $xy$-plane that creates the QDs is taken to be of the form
$$V_{\rm plane}(x,y)=\frac12m^*\left(\omega_x^2x^2+\omega_y^2y^2
\right)$$
where $\omega_x$ and $\omega_y$ are the confinement energies 
in the $x$- and $y$-directions respectively, corresponding to 
the oscillator lengths of $l_x = (\hbar /m^* \omega_x)^{-1/2}$ and 
$l_y = (\hbar /m^* \omega_y)^{-1/2}$. 
For $N$-electron system,  we also take into account the Coulomb 
interaction between the electrons
\[
{\cal H}_{\rm int}=\frac{e^2}{\epsilon}\sum_{i<j} 
\frac1{|\vec{r}_i-\vec{r}_j |}
\]
where $\epsilon $ is the background dielectric constant. We 
restrict the single electron basis by 18 lowest states and 
numerically obtain the eigenstates of the $N$-electron system 
with $N=2-9$. We analyze the shape, size and electron number
dependence on the the luminescence spectra of the QDCL. 

During the optical transition in the active region of a QDCL,
in the initial state (before optical emission) all electrons 
are in the second subband. In the final state (after 
optical emission) one electron is in the first subband, 
and all other electrons are in the second subband. The 
intensity of optical transitions is calculated from
\begin{eqnarray*}
&&{\cal I}_{if}(\omega)=\frac1Z\sum_{if}\delta(\omega-E_i+E_f)\\
&&\Bigg\vert\int\,\chi\Dagcomp_1(z)\,z\,
\chi\Dagcomp_2(z)\,dz\,\int \Phi_i^*(x\Dagcomp_1y\Dagcomp_1,
\cdots, x\Dagcomp_Ny\Dagcomp_N)\\
&&\times\Phi_f(x\Dagcomp_1y\Dagcomp_1,\cdots,x\Dagcomp_N
y\Dagcomp_N)\, dx\Dagcomp_1 dy\Dagcomp_1\cdots dx\Dagcomp_N
dy\Dagcomp_N \Bigg\vert^2\\
&&\times\exp(-\beta E_i)\\
\end{eqnarray*}
where $Z=\sum_i\,{\rm e}^{-\beta E_i}$ is the partition function 
and $\beta=1/kT$. In all our computation, we take $T=20$ K. 
To take into account disorder in the system we introduce the 
spreading of each emission line in the Lorentz form so that
the final intensity is
$${\cal I}(\omega)=\int d\omega_1 {\cal I}_{if}(\omega_1)
 \frac{\Delta}{\pi[\Delta ^2+(\omega-\omega_1)^2]}.$$
The parameter $\Delta$ in our calculation is taken to be 
$\Delta=2$ meV. Shell filling in a QD is characterized by sharp 
peaks in the current versus the gate voltage curves in 
single-electron tunneling spectroscopy. The corresponding 
addition energy shows sharp peaks for the electron numbers 
2, 6, 12, ... in the dot that could be interpreted as shell 
filling of the electronic states in the dot \cite{tarucha,shell}. 
Further, there are also weak peaks in addition energy for 
electron numbers 4 and 9 that are explained in terms of 
half-filled shell structure, in accordance with Hund's rule.
The addition energy is defined as the chemical potential 
difference $\Delta \mu_N=\mu(N+1)-\mu(N)$, where 
$\mu(N)=E(N)-E(N-1)$, and $E(N)$ is the ground state energy of 
the $N$ electron system.

\begin{figure}
\centerline{
\epsfxsize=3.0in
\epsfbox{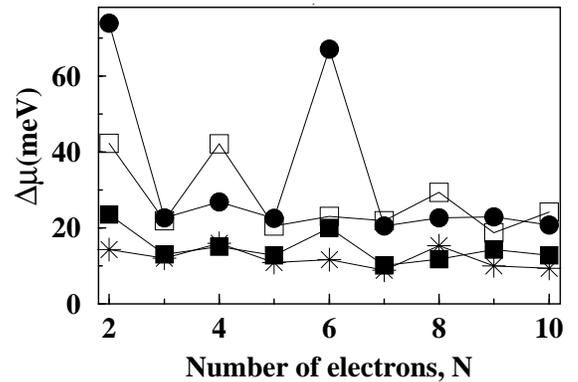}}
\vspace*{0.1in}
\protect\caption[addenergy]
{\sloppy{Addition energies of circular and elliptical quantum 
dots in the QC laser. Here the sumbols indicate: $\bullet$ for 
$(l_x=l_y=5$ nm), $\Box$ for $(l_x=5$ nm, $l_y=7$ nm),
$\bsqre$ for $(l_x=l_y=10$ nm),
$\star$ for $(l_x=10$ nm, $l_y=14$ nm),
}}
\label{addition}
\end{figure}

In the quantum-dot cascade structure we have also found  
peaks in addition energy as a function of the number of 
electrons. In Fig.~\ref{addition}, addition energies of circular 
and elliptical dots are presented. For small circular quantum dots 
($l_x =l_y =5$ nm) there are sharp peaks at $N=2$ and 6 as expected.
There are also peaks at $N=4$ and 9 where the total spin is
equal to 1 and 3/2, respectively. Because these 
peaks result from the interelectron interaction their strength
is much smaller than those at $N=2$ and $4$. The shell filling and
Hund's rule peaks are also observed in a larger dot 
($l_x =l_y =10$ nm). For elliptical dots there are no peaks 
corresponding to the Hund's rule ($N=9$). All peaks in this case 
corresponds to the shell filling at even values of electron number 
($N=2$, 4, 6 and 8). 

\begin{figure}
\centerline{
\epsfxsize=3.0in
\epsfbox{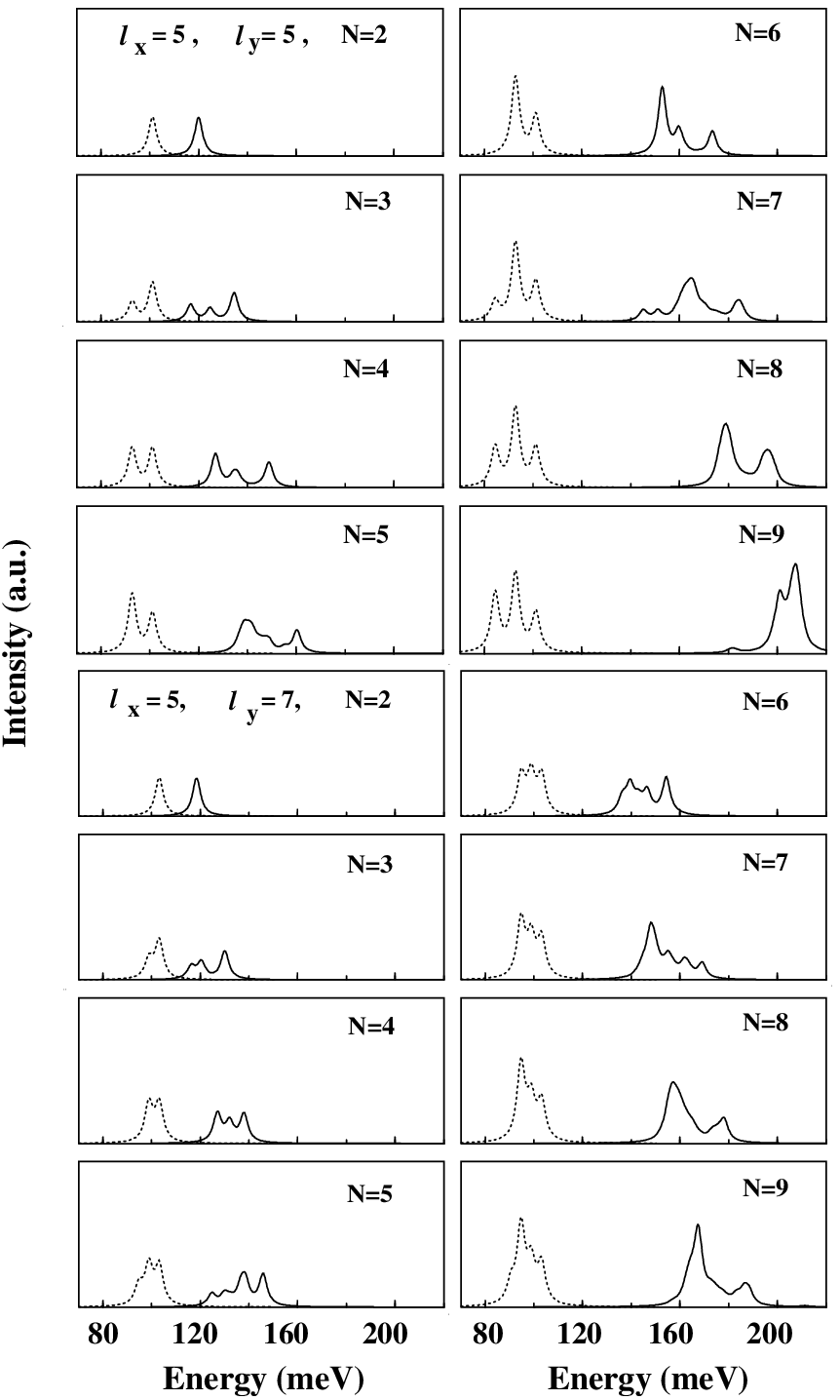}}
\vspace*{0.1in}
\protect\caption[luminescence5]
{\sloppy{Luminescence spectra of a quantum cascade laser with
circular $(l_x=l_y=5$ nm) and elliptical $(l_x=5, l_y=7$ nm)
quantum dots containing $N=2-9$ electrons, in the active region.
}}
\label{figlumin1}
\end{figure}

\begin{figure}
\centerline{
\epsfxsize=3.0in
\epsfbox{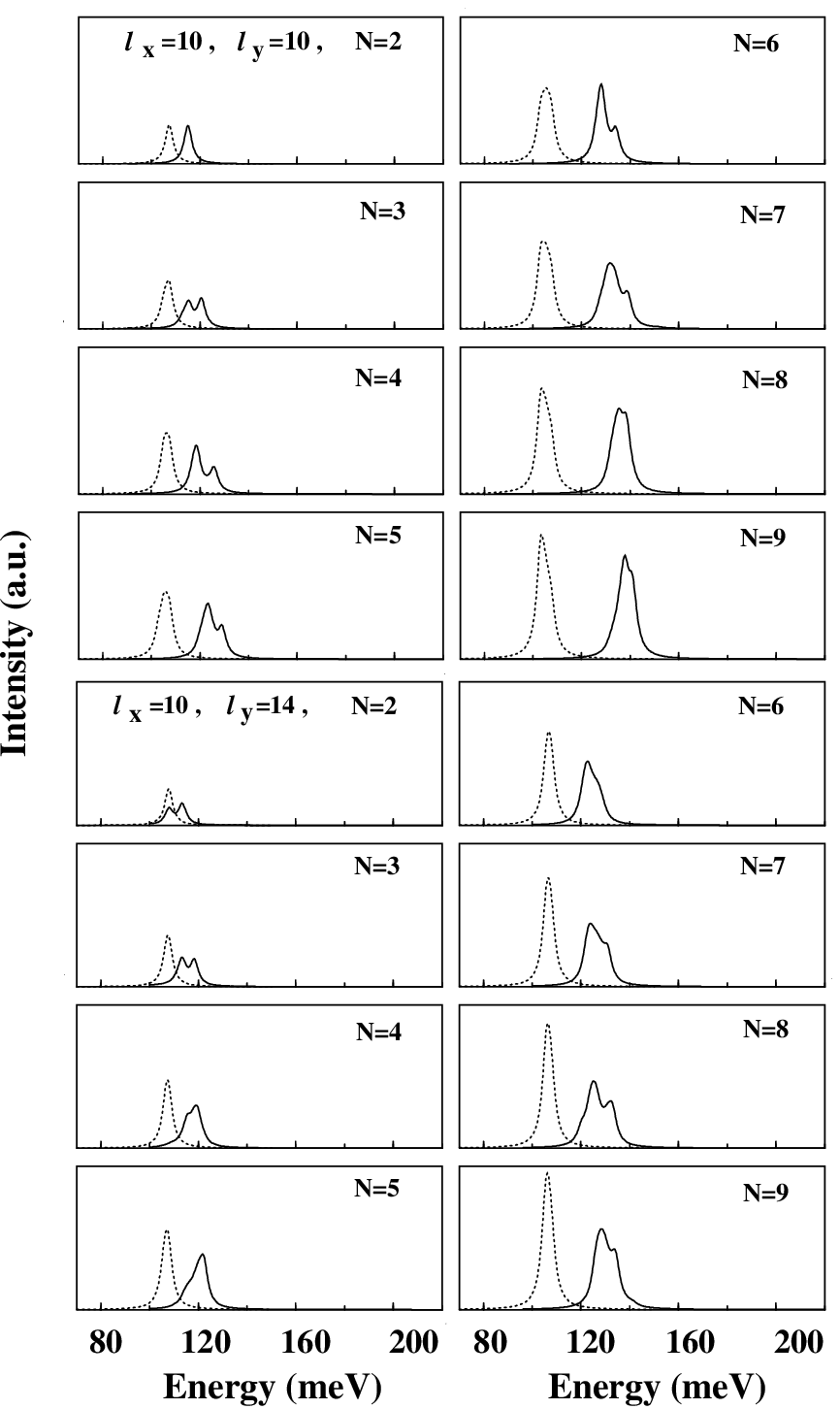}}
\vspace*{0.1in}
\protect\caption[luminescence10]
{\sloppy{Luminescence spectra of a quantum cascade laser with
circular $(l_x=l_y=10$ nm) and elliptical $(l_x=10, l_y=14$ nm)
quantum dots containing $N=2-9$ electrons, in the active region.
}}
\label{figlumin2}
\end{figure}

In Fig.~\ref{figlumin1} and Fig.~\ref{figlumin2}, we present the
luminescence spectra of our QC laser structures with circular 
and elliptic quantum dots with different number of electrons
in the active region. The emission 
spectra for non-interaction system are shown by dotted 
lines. For smaller quantum dots the emission lines of the
non-interacting system have some internal structure that is 
entirely due to the nonparabolicity. The nonparabolicity also 
gives the small redshift of emission line of non-interaction system. 
For large quantum dots there is single line for all number of 
electrons of non-interaction system, because non-parabolicity 
in this case becomes less important due to the smaller values of 
confinement energies $\omega_x$ and $\omega_y$. The 
electron-electron interaction results in a huge blueshift of 
the emission spectra compared to the results for
non-interacting electrons. The blueshift becomes smaller for 
larger quantum dots and also decreases for elliptic dots. This is
due to smaller interaction between the electrons when the spreading
of the electron wave function in $(x,y)$ plane becomes larger. 
The interaction between the electrons also changes the shape of 
the emission line. This becomes more important for smaller 
quantum dots (Fig.~\ref{figlumin1}). For larger quntum dots 
(Fig.~\ref{figlumin2}) the disorder makes the emission line almost 
single-peaked especially for large electron numbers when 
the interaction between the electrons becomes smaller. 

We can also see that for smaller quantum dots the change 
of the shape of the dots changes the shape of the emission 
line considerably. But for larger dots the shape of the emission 
line is less sensitive to the shape of the quantum dots and 
the line has almost the single peak both for circular and 
elliptic dots. Although the ground state of the electron system 
in the initial state obey the Hund's rule for circular dots 
(for $N=4$ and 9) we did not find any singularity in 
emission spectra due to the electron shell filling.  

The intersubband luminescence peaks of the QDC laser are rather
insensitive to the shell effect of the QDs in the active region.
For smaller (and circular) QDs, the peaks exhibit a huge
blueshift as the electron number is increased, that is entirely
due to the electron-electron interaction. The blueshift is smaller 
for larger and  non-circular dots. For large dots, shape of 
the emission line has weak dependence on the shape of 
quantum dots. We also present the addition energies of the QDCL.

We thank S. Blaser for helpful discussions on the QC lasers.
We also thank P. Fulde for his support and kind hospitality 
in Dresden.


\vfil

\end{document}